\documentclass[review,5p,twocolumn,times,10pt]{elsarticle}

\usepackage{graphicx}
\usepackage{fix2col}
\usepackage{amssymb}
\usepackage{amsthm}

\biboptions{numbers}

\usepackage{url}
\usepackage{booktabs}
\usepackage{tabularx}
\usepackage{multirow}   
\usepackage{txfonts}
\usepackage{bbding}     
\usepackage{textcomp} 
\usepackage[hang,small,nooneline]{caption}
\usepackage{fancyhdr}
\usepackage{fixltx2e}
\usepackage[hidelinks]{hyperref}
\usepackage{ulem}   
\newcolumntype{L}[1]{>{\raggedright\arraybackslash}p{#1}} 
\newcolumntype{C}[1]{>{\centering\arraybackslash}p{#1}} 
\newcolumntype{R}[1]{>{\raggedleft\arraybackslash}p{#1}} 

\newcolumntype{M}[1]{>{\centering\arraybackslash}m{#1}} 
\newcolumntype{N}[1]{>{\raggedright\arraybackslash}m{#1}} 

\newcolumntype{Y}{>{\centering\arraybackslash}X} 
	
\journal{Acta Astronautica}

\begin{document}

\begin{frontmatter}

\title{PANIC -- A surface science package for the \mbox{in situ} characterization \linebreak of a near-Earth asteroid}

\author[TUD]{Karsten~Schindler\corref{cor1}\fnref{fn1}}
\ead{schindler@mps.mpg.de}
\author[MIT]{Cristina~A.~Thomas\fnref{fn2}}
\ead{cristina.thomas@nau.edu}
\author[UND]{Vishnu~Reddy}
\ead{vishnu.kanupuru@und.nodak.edu}
\author[TUD]{Andreas~Weber}
\ead{andreas.weber@tu-dresden.de}
\author[TUD]{Stefan~Gruska}
\ead{stefan.gruska@mailbox.tu-dresden.de}
\author[TUD]{Stefanos~Fasoulas\fnref{fn3}}
\ead{fasoulas@irs.uni-stuttgart.de}

\cortext[cor1]{Corresponding author}

\address[TUD]{Technische Universit\"{a}t Dresden, Institute for Aerospace Engineering, 01062~Dresden,~Germany}
\address[MIT]{Massachusetts Institute of Technology, Department of Earth, Atmospheric and Planetary Sciences, 77~Massachusetts~Ave, Cambridge, MA~02139,~USA}
\address[UND]{University of North Dakota, Department of Space Studies, 4149~University~Ave~Stop~9008, Grand~Forks, ND~58202,~USA}

\fntext[fn1]{Current address: Max Planck Institute for Solar System Research, \hbox{Max-Planck-Stra{\ss}e 2}, 37191 Katlenburg-Lindau, Germany}
\fntext[fn2]{Current address: Northern Arizona University, Department of Physics and Astronomy, PO~Box~6010, Flagstaff, AZ~86011, USA}
\fntext[fn3]{Current address: Universit\"{a}t Stuttgart, Institute of Space Systems, \hbox{Pfaffenwaldring 31}, 70569 Stuttgart, Germany}

\begin{abstract}
This paper presents the results of a mission concept study for an autonomous micro-scale surface lander also referred to as PANIC -- the \textit{Pico Autonomous Near-Earth Asteroid In Situ Characterizer}. The lander is based on the shape of a regular tetrahedron with an edge length of 35\,cm, has a total mass of approximately 12\,kg and utilizes hopping as a locomotion mechanism in microgravity. PANIC houses four scientific instruments in its proposed baseline configuration which enable the in~situ characterization of an asteroid. It is carried by an interplanetary probe to its target and released to the surface after rendezvous. Detailed estimates of all critical subsystem parameters were derived to demonstrate the feasibility of this concept. The study illustrates that a small, simple landing element is a viable alternative to complex traditional lander concepts, adding a significant science return to any near-Earth asteroid (NEA) mission while meeting tight mass budget constraints. 
\end{abstract}

\begin{keyword}

asteroid \sep NEA \sep exploration \sep lander \sep in situ \sep small spacecraft

\PACS 96.25.-f \sep 96.25.Hs \sep 96.30.Ys \sep 96.25.Bd

\end{keyword}

\end{frontmatter}
\thispagestyle{fancy}

\section{Introduction}
\label{sec:introduction}

Aside from the limited compositional information gathered in situ after the landing of NEAR Shoemaker~\cite{Veverka2001,Evans2001} and during two touch-down maneuvers of Hayabusa~\cite{Yano2006}, no dedicated scientific study has been conducted on the surface of an asteroid yet. Two previous attempts of dedicated landers have been unsuccessful, although the primary spacecraft was the reason for failure in both cases: The PrOP-F lander~\cite{Sagdeev1987, Kemurdzhian1988}, a spherical ``hopper" with a diameter of $\approx$\,500\,mm, a mass of $\approx$\,50\,kg and a designed lifetime of 4\,h, was aimed to investigate the surface of the Martian moon Phobos (believed to be a captured asteroid resembling a C-type spectra~\cite{Lynch2007}) with eight instruments, resulting in a total payload mass of $\approx$\,8\,kg. However, contact with the carrier spacecraft was lost prior to its deployment. More recently, the \mbox{MINERVA} lander~\cite{Kubota2007}, the smallest planetary probe built to date, was released during an unexpected ascending maneuver of Hayabusa which resulted in a relative speed exceeding escape velocity of the target asteroid. Although the probe never reached the surface of Itokawa, telemetry data at least verified the lander operated nominally for $\approx$\,18\,h after deployment~\cite{Yoshimitsu2007}. MINERVA had a cylindrical shape with a length of 100\,mm, a diameter of 120\,mm and a weight of \textless\,600\,g. Its payload consisted of three cameras derived from commercial webcam modules and six temperature sensors.

Though a high level of miniaturization of space instruments and landers is currently feasible, the example of \mbox{MINERVA} illustrates that very small landers -- also referred to as surface science packages -- are very limited in their scientific capabilities. In contrast, complex traditional lander concepts with multiple instruments and objectives, such as the Philae lander on Rosetta~\cite{Bibring2007a}, exceed mass and cost budgets of low-cost small spacecraft exploration missions to \mbox{low-$\Delta$v} near-Earth asteroid (NEA) targets. 

The ten-week ``Small Satellite Summer Study Program (S4P)" held at NASA Ames Research Center in 2008 had the explicit goal to study options to use small cost-efficient spacecraft for future missions to NEAs while being aware of their limitations. One result of the program is the Discovery-class mission concept ``Didymos Explorer (DEx)"~\cite{Cook2008,Rozitis2009} intended to investigate the binary asteroid \mbox{(65803)~Didymos}, a potentially hazardous NEA with a spectral class not yet visited. It became clear that \mbox{in situ} surface investigations are an essential contribution to future NEA exploration missions as it is impossible to study micro-scale surface features remotely, making a lander also a key element of DEx. Still, it remained unclear if a lander can fit into the mass limitations of a small spacecraft. As a lander is a fully independent, specialized spacecraft on its own, having fundamentally different requirements and constraints than the orbiter, a separate study was necessary to derive reliable estimates to answer this question.

This lead to the concept study of a small lander called ``PANIC", the ``Pico Autonomous Near-Earth Asteroid In Situ Characterizer", which has been conducted by a second group in parallel to DEx and also continued after the program. The study's motivation was to determine the feasibility of a micro-scale lander concept that will provide a balance between spacecraft size and science return. A specific design concept was developed based on information about currently available technologies and hardware estimates. The results of this study will be summarized in this paper.

\section {Mission Objectives \& Payload}
\label{sec:objectives}

Remotely acquired reflectance spectra of asteroids are likely altered by processes of ``space weathering" (e.g. lower albedo, slope-reddening; see e.g.~\cite{Clark2002} for details) which hinders their interpretation regarding mineralogical composition and possible links to known meteorites. So far, space weathering effects could only be studied directly using returned lunar soil samples. To be able to better constrain the composition of NEAs, it is a key scientific interest to study the surface properties and structure on the micro-scale in situ in support of or as an alternative to a sample return. The following main scientific objectives have been defined for a lander mission by the study team as a starting point for further investigations (``straw man proposal"): 
\begin{enumerate}
\item{Characterize the bulk composition and geochemistry. Establish a link between the target body and a meteorite class already known.}
\item{Investigate the particle size distribution on the surface.}
\item{Study and constrain the effects of space weathering regarding changes in the optical characteristics.}
\end{enumerate}
Secondary mission goals are:
\begin{enumerate}[(a)]
\item{Demonstrate mobility in a microgravity environment through hopping for the first time.}
\item{Study surface diversity through measurements at multiple locations.}
\item{Demonstrate an advanced level of miniaturization in planetary probe design.}
\end{enumerate}

A comprehensive payload survey revealed a number of highly miniaturized instruments which had already been developed for previous missions or are currently in development. Table~\ref{tab:Instruments} summarizes the selected payload, consisting of four instruments with a total payload mass of about~1.4\,kg and the related science traceability.

\begin{table*}[htbp]
\footnotesize
\centering
\caption{Selected baseline instrumentation in connection with PANIC's main scientific objectives: (1) Characterize the bulk composition and geochemistry. (2) Investigate the particle size distribution. (3) Study and constrain the effects of space weathering.}
\label{tab:Instruments}

\begin{tabularx}{\textwidth}[t]{>{\raggedright\arraybackslash}X N{4.2 cm} ccM{3.2cm}ccc} \toprule
Instrument & Component & \renewcommand{\multirowsetup}{\centering} \multirow{2}{1.2cm}{Mass \linebreak \lbrack{g}\rbrack} & \renewcommand{\multirowsetup}{\centering} \multirow{2}{1.2cm} {Margin \linebreak \lbrack{g}\rbrack} & Heritage & \multicolumn{3}{c}{Objective} \\ \cmidrule{6-8}
& & & & & 1 & 2 & 3 \\ \midrule

\multirow{2}{3.8cm}[-3px]{Alpha Particle \mbox{X-ray} Spectrometer (APXS)} & Sensor Head & 115 & 12 & \renewcommand{\multirowsetup}{\centering} \multirow{2}{3cm}[-3px]{MER~\cite{Rieder2003}, Philae~\cite{Klingelhoefer2007}, Nanokhod~\cite{Schiele2008}} & \renewcommand{\multirowsetup}{\centering} \multirow{2}{*}[-3px]{\Checkmark} & & \\ \cmidrule {2-4} 
& Electronics & 120 & 12 & & & & \\ \midrule

\multirow{5}{3.8cm}[1px]{Near-Infrared Spectrometer (NIRS)} & Sensor Head 	& 80 & 16 & \renewcommand{\multirowsetup}{\centering} \multirow{5}{2.5cm}[1px]{R\&D \linebreak (Examples: MUSES-CN~\cite{Jones2000},~\cite{Luethi2008},~\cite{Leroi2009})} & \renewcommand{\multirowsetup}{\centering} \multirow{5}{*}[1px]{\Checkmark} & & \renewcommand{\multirowsetup}{\centering} \multirow{5}{*}[1px]{\Checkmark} \\ \cmidrule {2-4} 
& Electronics 								& 120 & 24 & & & & \\ \cmidrule {2-4} 
& Illumination, TEC, \linebreak Mirror~\& Lens Assemblies 			& 300 & 90 & & & & \\ \midrule

Microscopic Imager (MIC) & Camera Module incl. \linebreak Optics, Illumination and Electronics & 165 & 17 & Beagle 2~\cite{Thomas2004, Website_Bern_MIC} & & \Checkmark & \Checkmark \\ \midrule

Stereo Camera (SC) & Camera Modules (2 pcs) \linebreak incl. Optics and Electronics & 280 & 28 & Beagle 2~\cite{Griffiths2005}, Philae~\cite{Bibring2007b}, PROBA, \dots (Key Ref.:~\cite{Beauvivre2008}) & & \Checkmark & \\ \midrule

Total incl. Margins & & \multicolumn{2}{c}{1379} & & & & \\ \bottomrule
\end{tabularx}                       
\end{table*} 

The Alpha Particle X-ray Spectrometer (APXS)~\cite{Rieder2003,Schiele2008} will be used to directly determine elemental abundances at the landing site. The Microscopic Imager (MIC) having a spatial resolution of 6\,\textmu \mbox{m/pixel}~\cite{Thomas2004, Website_Bern_MIC} will investigate the grain size distribution and search for evidence of rims formed by nano-phase iron on individual grains, a known product of space weathering~\cite{Pieters2000}. A precision linear stage is required to move the APXS to its specified working distance and the MIC to a sequence of focus positions.

A Near-Infrared Spectrometer (NIRS)~\cite{Jones2000} will be used to study the mineralogy and optical properties at wavelengths of 0.8~--~2.5\,\textmu m. The mobility of the lander will allow for a variety of surface spectra taken at different locations and under different environmental conditions. These measurements will improve scientists' understanding of optical surface effects such as space weathering, particle size and surface temperature. The results have the potential to constrain the influence of surface effects on NIR spectroscopy.

The Stereo Camera (SC) system~\cite{Griffiths2005, Bibring2007b} will allow imaging of the surrounding terrain in one direction from the lander using its wide-angular optics (\textgreater\,60\textdegree) and measure the distance and size of geological surface features. Acquired imagery after each successive hop will provide a random survey of the surface. The acquired images are also vital for public outreach purposes, since the picture could be the first taken from an asteroid's surface. The stereo approach provides redundancy -- in case one of both cameras fails, the remaining data will still impart an impression of the landing site. 

\section {Basic Approach}
\label{sec:approach}

The study assumes a target body with a size and rotational period similar to (25143)~Itokawa (\mbox{$\approx$\,550\,m~\texttimes~300\,m~\texttimes~225\,m; P~=~12.1\,h}) -- an object expected to be representative in size, shape and physical properties for a major fraction of the NEA population~\cite{Fujiwara2006}. This implies a surface gravity on the order of 10\,\textmu g~\cite{Scheeres2010} -- comparable in its magnitude to the remaining accelerations aboard the International Space Station or sounding rockets~\cite{ZARM} -- and escape velocities on the order of \mbox{10~--~20\,cm/s~\cite{Fujiwara2006}}. The lander needs to be carried by a primary spacecraft which is heading for a rendezvous orbit at the asteroid and releases the lander during a low-altitude maneuver in close proximity. In a matter analogous to both the previously mentioned and several other proposed small body exploration landers (e.g.~\cite{Jones2000, Cottingham2009}), PANIC will employ an uncontrolled passive free fall and a hard landing dependent on suitable deployment from its primary orbiting spacecraft. Additionally, the lander will have self-righting capabilities to enable an advanced payload. This allows to abandon the attitude control subsystem and save the related subsystem mass.

Mobility to study surface diversity is a mission requirement. Still, the mission will be counted as a success if one complete set of science data is gathered at the initial landing site and successfully transmitted to the orbiting spacecraft. The lander will aim to acquire data at a minimum of two different sites in order to sample surface diversity. Considering the necessary durations of data collection and uplink for a minimum of two sites, the lifetime of the lander has to be on the order of \mbox{24~--~36\,h}.

The study's target mass of $\approx$\,10\,kg led to the development of a ``pod lander" concept -- a lander, which can place itself in a defined orientation by means of external actuators such as petals or rods. Although penetrators enable subsurface access in microgravity, they are limited to single point measurements and have not been successfully demonstrated on previous interplanetary missions (see failures on Mars 96, DS-2~\cite{Ball2007}). Since porosity and mechanical resistance of the upper surface layers of asteroids are poorly understood, using a penetrator concept remains a high risk mission. Conventional rovers are not an appropriate solution for NEA exploration due to the microgravity environment. The canceled \mbox{MUSES-CN} micro hybrid rover-hopper~\cite{Jones2000,Jones1999} is a prime example of the limitations of rovers on NEA surfaces. It was estimated that surface dynamics would limit its maximum speed to $\approx$\,1\,mm/s.

From all geometries considered by the study team, a tetrahedral shape appears to be the most favorable for a pod lander. A sphere has been discarded as it might have severe difficulties settling on a surface full of slopes and depressions within reasonable time scales. A tetrahedron has the lowest number of faces of all polyhedrons and hence requires only three actuators for self-righting. The simple straight geometries of the design appear to be beneficial in terms of fabrication and utilization of volume. The synchronous deployment of radially symmetrically arranged petals is advantageous as the applied torques cancel each other out. A tip over of the lander during the self-righting process on a rough, boulder-rich or sloped surface can be intercepted by the petals in every direction. Moreover, PANIC can directly use its petals to induce a hop by quickly pushing them into the soil, while limited control on the direction of the hop could be implemented through slightly different petal actuation. In contrast, a discoidal design -- a pod lander with only one lid as on Beagle~2~\cite{EuropeanSpaceAgency2004}, which is able to flip over -- does not allow any control over hopping direction as it has only one actuated petal.

Although imprecise, this hopping mechanism provides a robust mobility system which can overcome the perils of rough asteroid surfaces and enables the lander to travel large distances on rather short time scales. The mobility of the lander allows for a representative surface sampling within its short lifetime, which is limited by the capacity of its non-rechargeable primary batteries. 

\section {Mission Sequence}
\label{sec:ops}

In a manner analogous to the Hayabusa mission, this study assumes the lander is released during a low altitude maneuver of several tens of meters at a relative velocity close to zero. This leads to an impact velocity on the order of 10\,cm/s, a value which must be below the local escape velocity to safeguard the lander against loss following a rebound from a very hard surface. It is further assumed that the orbiter will ascend to an inertial sub-solar position (between the Sun and the asteroid, \mbox{5~--~10\,km} altitude) to hover over the asteroid after release of the lander. Therefore, data transmission in the nominal mission sequence is limited to daytime periods only. A third assumption in order to limit the parameter space for this study shall be that day and night have the same length (about 6\,h). Of course, the actual hours of daylight depend greatly on the orientation of the asteroid's rotation axis and latitude of the landing site.

For landing and hopping simulations, the initial simplification in our model has been a triaxial ellipsoid asteroid with constant density whose gravitational field is represented using spherical harmonic expansions (see e.g.~\cite{Hu2002}). However, we eventually came to the conclusion that spherical harmonic series only converge outside the circumscribing sphere of the ellipsoid (the so-called ``Brillouin sphere"). Severe divergence can happen inside of this sphere, especially the deeper the test point mass is inside of it. Therefore, this approach only works for a simplified orbit analysis of the primary spacecraft, but not to describe the movement of a lander very close to the surface. A much better approach for the analysis of the lander dynamics is the use of ellipsoidal harmonic expansions which converge outside the body's circumscribing ``Brillouin ellipsoid" that is congruent to the hypothetical ellipsoidal asteroid (see~\cite{Garmier2001,Dechambre2002} for more details). An alternative approach would be to use a polyhedron~\cite{Werner1997} or a finite element shape model (e.g. consisting of cubes or spheres)~\cite{Park2010} to calculate the asteroid's gravity field.

To work around the divergence errors of spherical harmonic expansions, it was necessary to further simplify the asteroid to a sphere with an equivalent radius derived from Itokawa's volume. These simplified simulations still show that the lander will need considerable time to settle after the initial landing or following a hop. Depending on the surface's coefficient of restitution (literature estimates: \mbox{0.5~--~0.8}~\cite{Korycansky2006}; measurement on Itokawa: 0.84~\cite{Yano2006}), coefficient of friction (Itokawa: \textgreater\,0.8~\cite{Yano2006}) and remaining relative velocity after release (\mbox{assumption: -5~--~5\,cm/s}), settling times on the order of \mbox{1.2~--~6\,h} after the release from the orbiter and \mbox{20~--~60\,min} after a hop have been estimated. The lander still consumes power during these time periods as it must be able to detect its own settlement and daylight phase.

The on-board science instruments (with the exception of the SC) would benefit from night time operations. The MIC and NIRS can be calibrated to implemented artificial light sources which illuminate their targets. This allows a more reliable interpretation of the acquired images and spectra, which would be at risk to be contaminated by solar stray light during day time. The NIRS detector has to be cooled by a thermo electrical cooler (TEC) to sufficiently low temperatures. The APXS is required to operate at low temperatures in order to keep the detector noise at an acceptable level. Keeping the limited battery capacity of the lander in mind, night time operations are a requirement to allow for valuable scientific investigations as it is virtually impossible to operate -- and therefore cool -- the NIRS and APXS during daytime hours.

\begin{figure*}[htbp]
\centering
\includegraphics[width=\textwidth]{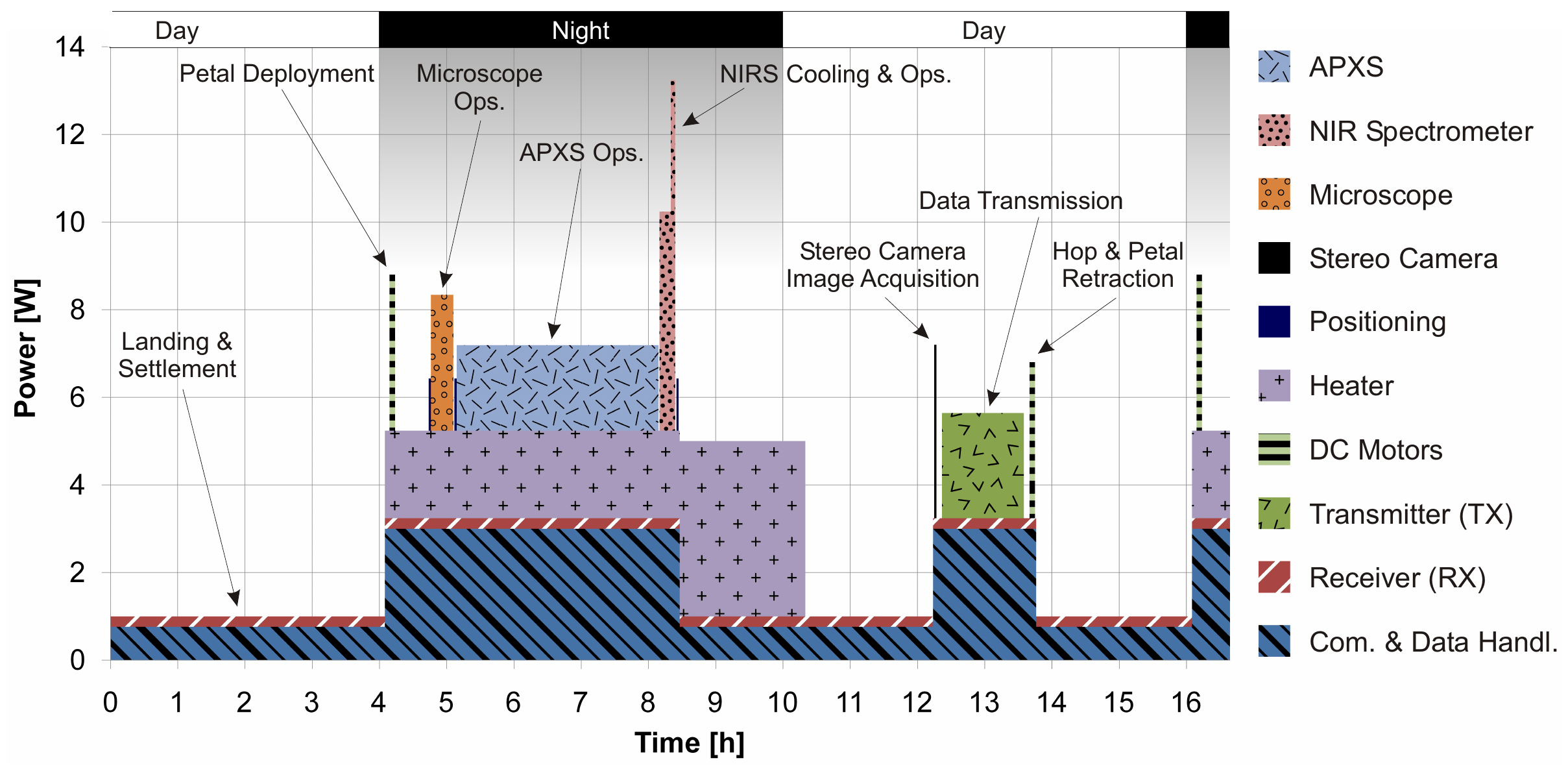}
\caption{The diagram illustrates an exemplary mission sequence for one asteroid day after release from the orbiter assuming payload operations are limited to night time periods. The target's rotational period shall be 12\,h while day and night shall have the same length. Dyed areas mark the diurnal time without sunlight. The lander is released during a low-altitude maneuver and settles before sunset. The beginning of the night initiates self-righting and science operations. The collected data is stored in a mass memory, as no communication link can be established during night time to the orbiter hovering in an inertial position between the Sun and the asteroid. Apart from the acquisition of a pair of stereo images and data uplink, e.g. triggered by a telecommand relayed by the orbiter, the lander is in stand-by mode during the remaining part of the day. The calculated energy consumption for the acquisition of two data sets as listed in Table~\ref{tab:DataVolume} during two asteroid days, also considering settling and idle times as illustrated, is 108\,Wh within a life-time of 26\,h.}%
\label{fig:missiontimeline}
\end{figure*}

As discussed later in Section~\ref{sec:subsystems}, night time operations would also have a positive impact on thermal control. The heat dissipated by the electronics would help to keep the lander compartment warm. This reduces the necessary heat emission of an active heater which ensures that the lander's temperature does not fall below critical values for operation and storage of the implemented hardware. As less energy is consumed for heating, a second hop and third instrument cycle are feasible with the available battery power at battery temperatures above 0\,\textdegree C. Figure~\ref{fig:missiontimeline} illustrates an example mission sequence for a scenario where instrument operations (except for the SC) are limited to night. 

The lander approaches the target and settles on the surface in an arbitrary orientation after a number of bounces. In this scenario, the side petals are deployed when the craft is at rest and nightfall has occurred, pushing the lander into an upright position. Once the lander is in the appropriate configuration, the MIC, APXS and NIRS will subsequently acquire data. A stereo image is taken following sunrise. Data transmission is triggered by the orbiter sending a specific command. After the uplink has been successfully finished, the lander relocates through hopping, retracts its petals on its initial ballistic trajectory and comes again to a rest on the surface after several bounces. It remains in stand-by until nightfall and repeats its data acquisition, uplink and hopping cycle until the batteries are drained.

Table~\ref{tab:Power} introduces the average power requirements of all components and subsystems included in the lander's power budget and operations schedule. An energy of $\approx$\,30.25\,Wh (incl.~10\%~safety margin) and a time period of $\approx$\,5\,h are necessary to obtain the measurements as specified in Table~\ref{tab:DataVolume}, to transmit the acquired data to the orbiter and to keep the command and data handling subsystem (C\&DH) and command receiver (RX) switched on during instrument data acquisition and uplink. The minimum duration of instrument data acquisition is mainly driven by the long integration time of the \mbox{APXS (2~--~3\,h~\cite{Rieder2003,Klingelhoefer2007})}. However, current developments indicate a significant reduction in integration time for future APXS instruments thanks to a significant increase in sensitivity~\cite{RichterTalk2010}. 

\begin{table}[htbp]
\footnotesize 	
\centering  	
\caption{Estimated power requirements of the lander's subsystem components.}

\begin{tabularx}{\columnwidth}{>{\raggedright\arraybackslash}X C{2cm} C{1.2cm}} \toprule  
Component & Average Power~\lbrack{W}\rbrack & Margin \lbrack{W}\rbrack\\ \midrule
APXS & 1.1 & 0.1 \\ 
Microscopic Imager (MIC) & 2.7 & 0.3 \\ 
NIRS - Instrument & 3 & 0.9 \\
NIRS - Thermo Electrical Cooler & 5 & 1\\
Stereo Camera (SC)	& 3.6 & 0.4 \\
Petal Motors	& 3.2 & 0.3 \\ 
Precision Stage Motor 	& 1.1 & 0.1 \\ 
Command \& Data Handling Unit & 2.5 & 0.5 \\
S-Band Transmitter (TX) & 2.0 & 0.4 \\ 
UHF Receiver (RX) & 0.2 & 0.05 \\ \bottomrule

\end{tabularx} 
\label{tab:Power}                      
\end{table}   

\begin{table*}[htbp] 
\small
\centering
\caption{Data volume accumulated by the payload.} 
\label{tab:DataVolume}
\begin{tabularx}{\textwidth}{>{\raggedright\arraybackslash}X Y Y Y Y Y} \toprule
Instrument & Resolution \linebreak \lbrack{pixels}\rbrack & ADC Resolution \linebreak \lbrack{bit}\rbrack & No. of \linebreak Measurements \linebreak per Data Set & Assumed \linebreak Compression \linebreak Ratio & Transmitted \linebreak Data Volume \linebreak \lbrack{kbyte}\rbrack \\ \midrule
APXS & n/a & n/a & 1 & none & 10 \\
NIRS & 256~\texttimes~1 & 12 bit & 10 & none & 4 \\
MIC & 1024~\texttimes~1024 & 10 bit & 60 & 10:1 (LZW) & 7680 \\
SC & 1024~\texttimes~1024 & 10 bit & 2 & 1.8:1 (Wavelet) & 1422 \\ \midrule
\multicolumn{5}{l}{Total} & 9116 \\
\multicolumn{5}{l}{Total +10\% Margin} & 10028 \\ \bottomrule
\end{tabularx} 
\end{table*}                      

The lander is in stand-by for the remaining time of the asteroid day. Accounting for the power which is required by an active thermal control subsystem during night time and by the on-board electronics running in low-power mode during bounces, settling and idle times, the acquisition of two data sets consumes about 108\,Wh.

Table~\ref{tab:DataVolume} shows the amount of data accumulated by the instruments which adds to about 10\,MB. The Microscopic Imager is likely to produce the largest amount of data, since many images ($\approx$\,60) are necessary due to the very short focal length and resulting depth of field ($\approx$\,40\,\textmu m, as estimated for Beagle~2~\cite{Thomas2004}). This results from the fact that the exact focus position is unknown a priori. Different parts inside the field of view will have different focus positions due to surface roughness. Therefore, many images at different focal positions must be taken in order to guarantee that all aspects of the field are in focus within some subset of the dataset. A Lempel-Ziv-Welch~(LZW) compression algorithm can effectively reduce the data amount since unfocused pictures compress at a rate exceeding \mbox{$1:40$}~\cite{Thomas2004}. The study assumes an average compression rate of \mbox{$1:10$} for data from the Microscopic Imager.

The pictures provided by the Stereo Camera shall be compressed using Wavelet compression at a rate of \mbox{$\approx$\,1.8:1} (value taken from ROLIS~\cite{Mottola2007, DLR_ROLIS}). The APXS and NIRS data will be left unaltered. In order to increase the signal-to-noise ratio (SNR) of the NIRS data, ten measurements shall be acquired. Instrument calibrations should be done using the orbiter's power supply before the lander is released.

\section {General Design and Configuration}
\label{sec:design}
\begin{figure}[htbp]
\centering
\includegraphics[width=\linewidth]{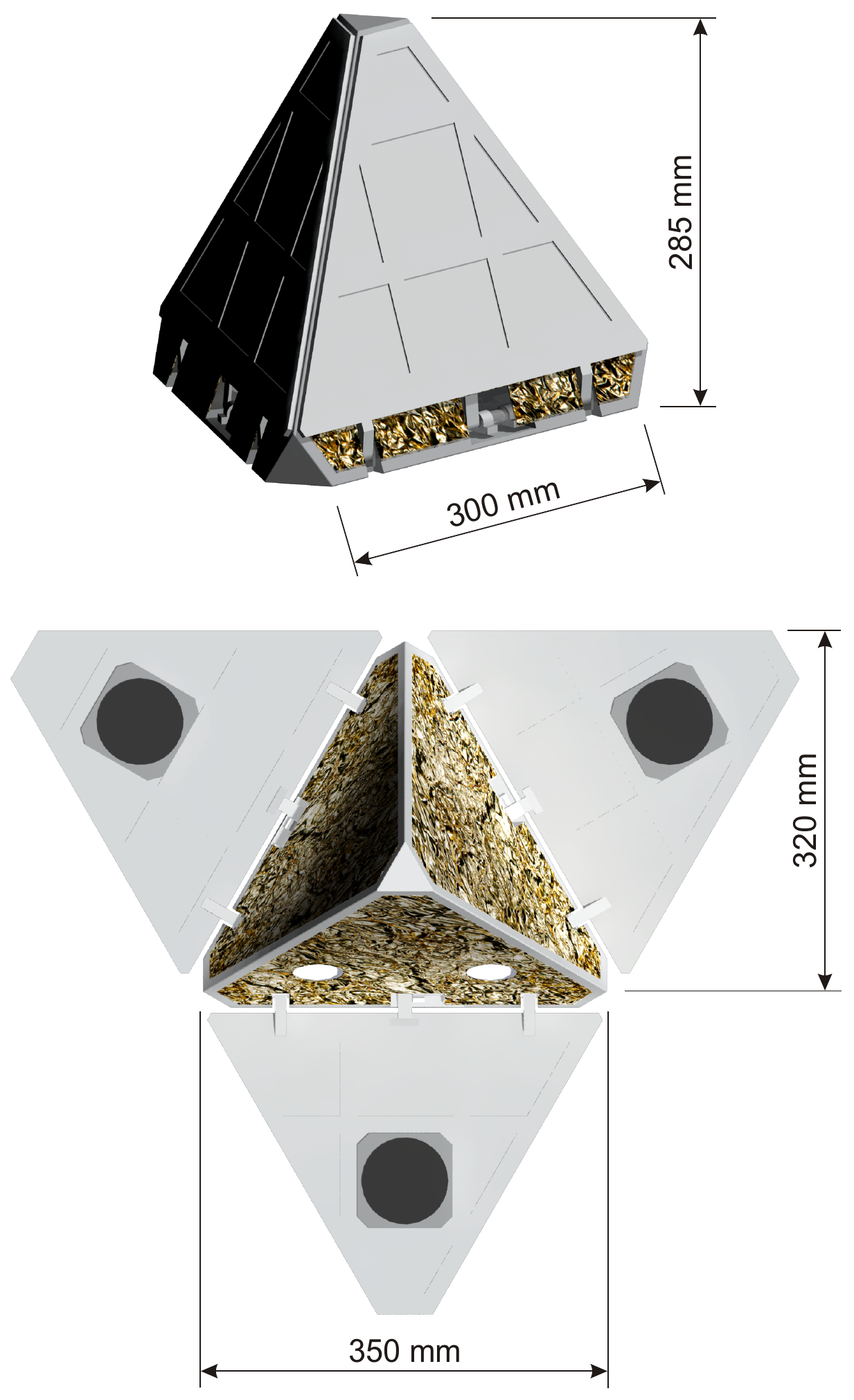}
\caption{The lander's design as a proof-of-concept \mbox{CAD-model}. The lander's external dimensions can be inscribed into a box sized  350~\texttimes~320~\texttimes~285\,mm\textsuperscript{3}.}
\label{fig:landerdesign}
\end{figure}

\begin{figure*}[htbp]
\centering
\includegraphics[width=\textwidth]{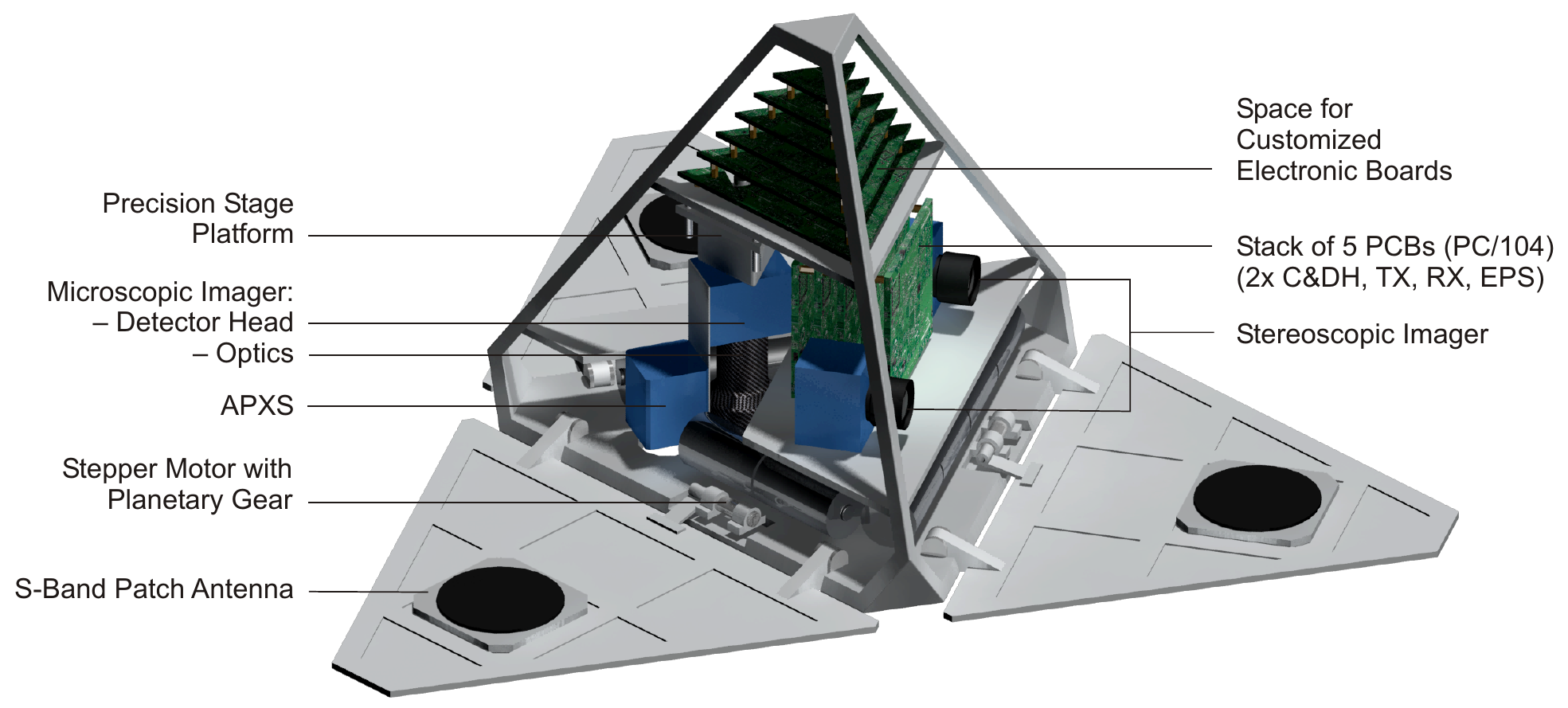}
\caption{Suggested arrangement of components in the lander's interior as a proof-of-concept CAD-model. Aluminum PCB cases are masked out for clarity.}
\label{fig:components}
\end{figure*}

\begin{figure*}[htbp]
\centering
\includegraphics[width=\textwidth]{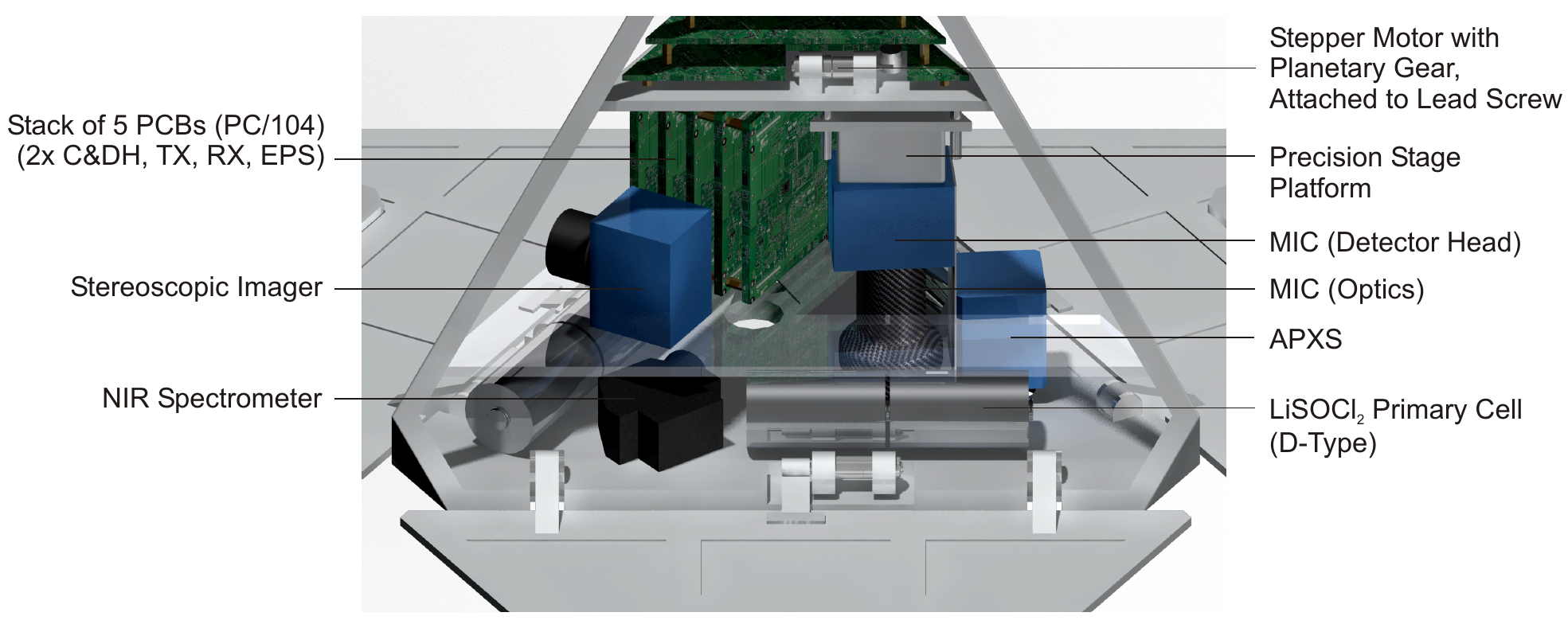}
\caption{Close-up view of the landers inner configuration and payload. Aluminum PCB cases are masked out for clarity.}%
\label{fig:components_closeup}
\end{figure*}

A proof-of-concept model was created to determine the necessary size and resulting weight. PANIC's general design is illustrated in Figure~\ref{fig:landerdesign}, while a suggested inner configuration is displayed in Figure~\ref{fig:components} and~\ref{fig:components_closeup}. The lander's structure offers a total volume of $\approx$\,6.75\,dm\textsuperscript{3}. In its stowed configuration, the entire unit can be circumscribed by a box of $\approx$\,350~\texttimes~320~\texttimes~285\,mm\textsuperscript{3}. 

Printed circuit boards (PCBs) required by the instruments and subsystems shall form two distinct packages: One stack consisting of five PCBs (1x~Receiver -- RX, 1x~Transmitter -- TX, 1x~Electrical Power Subsystem -- EPS, 2x~Command and Data Handling Subsystem -- C\&DH), and a customized pyramidal board stack in the lander's tip consisting of six triangular PCBs of decreasing area. The pyramidal stack provides space for the APXS and NIRS front end electronics, further payload related PCBs (e.g. stepper motor drivers, accelerometer) and additional subsystem related electronics. Both PCB stacks are encased in aluminum boxes (not shown in figures). The front end electronics of the MIC and SC are implemented into the integrated imager heads~\cite{Beauvivre2008}. The batteries are mounted at the lander's baseplate causing a lower center of gravity. The NIRS, also located at the lander's base, points at the surface through a window via a tilted mirror assembly. The MIC and APXS are attached to a translation stage to move them into focus.

\section {Subsystems}
\label{sec:subsystems}

Nano satellite missions in recent years have demonstrated the use of ruggedized industrial hardware designed for harsh environments or automotive applications in low Earth orbits (LEOs)~\cite{Helvajian2009}. Previous interplanetary technology demonstrator probes such as Sojourner~\cite{Matijevic1997} and MINERVA~\cite{Kubota2007} have successfully shown that a design based on wisely selected commercial-of-the-shelf (COTS) parts, integrated in custom solutions, can meet the requirements of the mission. The PC/104 standard, a convenient form factor for modular circuit boards with a geometry of 3.55{$''$}~\texttimes~3.775{$''$} (90.2\,mm~\texttimes~95.9\,mm), appears to prevail for nano satellite applications as it offers easy integration, stacking and mounting capabilities~\cite{ISIS_Structure, Kalman2008, Ubbels2009}. It is important to note that almost all adequate small components are not yet rated for deep space missions and still require proper qualification and test before use. However, their documented flight heritage in LEO can be an argument for justifying their selection, especially for a mission with an extremely limited lifetime. 

\paragraph{Structure and Mechanisms Subsystem (SMS)} 

The outer tetrahedral structure and petals will be manufactured from an aluminum alloy. Intermediate floor panels inside the lander shall be made out of sandwich plates, consisting of an aluminum honeycomb core with carbon fiber reinforced plastic (CRFP) sheets on both sides. Battery casings shall be exclusively made of CFRP. Each side petal and the MIC~/~APXS precision stage is driven by a stepper motor which is translated by a planetary gear head.

The structure should be completely encased by crushable material which acts as a damper to safeguard PANIC against leaving the asteroid's sphere of influence due to an excessive rebound velocity and to reduce settling time on the surface. The entire surface of the structure (having a total area of $\approx$\,0.27\,m\textsuperscript{2}) can be covered with foam material protecting the lander during multiple bounces in arbitrary orientations before coming to a final stop. The total mass of an applicable foam material~\cite{Rohacell2009} would be less than 50\,g (assumed thickness: 5\,mm). If the required amount of glue is also taken into account, the impact on the mass budget would be an acceptable additional mass of about 100\,g.

Due to the intended use of COTS components, radiation shielding is a main design driver. Although the lifetime of the lander on the surface will be very short, the lander will likely spend many years in space and must have sufficient radiation tolerance for this environment. Due to the absence of a shielding atmosphere, the radiation environment on the surface of an asteroid is considerably higher than on other bodies such as Titan, Venus and Mars. It is assumed that airless bodies also emit secondary radiation from their surface as a result of secondary particle interactions. The received radiation flux on the surface is comparable in value to that received by the orbiter en route to the asteroid~\cite{Ball2007}.

Assuming low radiation tolerance of COTS components (usually on the order of \mbox{2~--~10\,krad}~\cite{Tribble2003}) and a \mbox{30~--~36~month} long cruise phase (as originally planned on NEAR and realized on Hayabusa), a shielding equivalent of an aluminum wall shell with a thickness between \mbox{4.6~--~5.8\,mm} is required for an annual radiation dose of \mbox{0.6~--~3.3\,krad}. Since the lander already has a protective shield in its stowed configuration due to its base plate and closed side petals (which in spite of the lander's small kinetic energy on impact need to have sufficient strength to withstand impact and bouncing on the surface), a wall thickness of 4\,mm of the PCB's aluminum cases appears to be conservative and sufficient to safeguard all on-board electronics.

\paragraph{Thermal Control Subystem (TCS)}

The thermal conditions on an asteroid for a landing element are mostly unknown. No direct surface temperature measurement has been conducted on an asteroid so far. Only the Hayabusa mission was able to determine a surface temperature of $\approx$\,310\,K close to the landing site on Itokawa via an indirect radiative measurement (equilibrium temperature of the radiator panel during hovering in proximity; solar distance $\approx$\,1\,AU~\cite{Yano2006}).

\begin{figure*}[htbp]
\centering
\includegraphics[width=\textwidth]{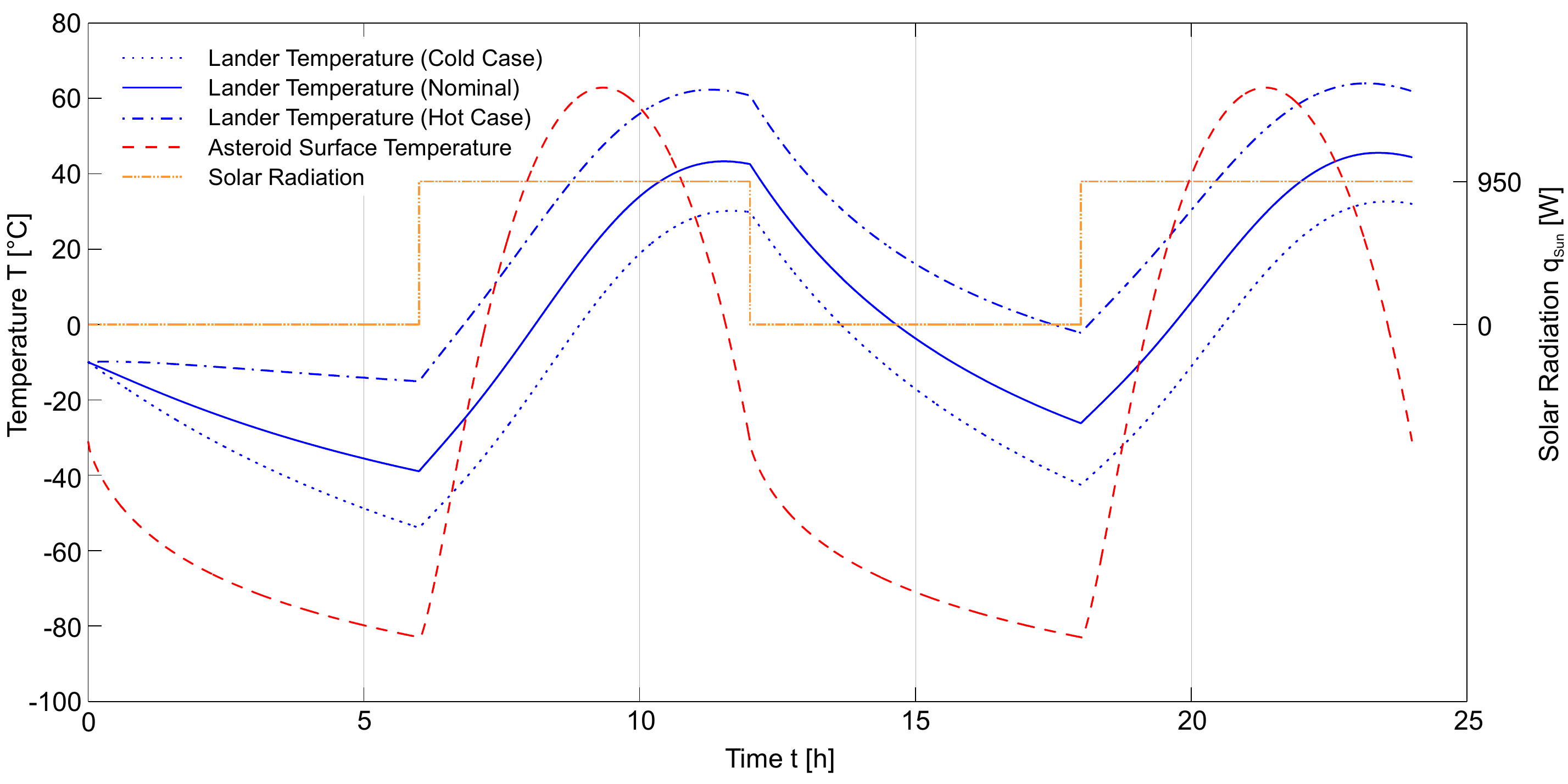}
\caption{Expected temperature profile of PANIC during the day and night phase on the asteroid's surface. Calculations assume an Itokawa-like target with a thermal inertia of 200\,J\,m\textsuperscript{-2}\,s\textsuperscript{-0.5}\,K\textsuperscript{-1}, an albedo of 0.14, an emissivity of 0.9 and a solar distance of 1.2 AU.}
\label{fig:thermal}
\end{figure*}

A simplified thermo-physical model (TPM, see~\cite{Harris2002,Mueller2006} and citations therein) has been used to derive the temperature profile in the course of an asteroid day. Surface temperature is mainly a function of the regolith's thermal inertia, albedo and solar distance. Figure~\ref{fig:thermal} illustrates one representative case of a variety of studied scenarios, which has been derived using the following boundary conditions: I) A thermal inertia of 200\,J\,m\textsuperscript{-2}\,s\textsuperscript{-0.5}\,K\textsuperscript{-1} (average value for the NEA population estimated by \mbox{Delbo et al.~\cite{Delbo2007}}); II) an albedo of 0.14 (average value for the NEA population estimated by \mbox{Stuart~\cite{Stuart2003}}); III) an emissivity of 0.9 (\mbox{Delbo et al.~\cite{Delbo2007}}); IV) a solar distance of 1.2 AU during operations.

The resulting surface temperature from this model varies between $\approx$\,-83\,\textdegree C and +63\,\textdegree C. Higher thermal inertia (\mbox{cf. Itokawa: 750\,J\,m\textsuperscript{-2}\,s\textsuperscript{-0.5}\,K\textsuperscript{-1}~\cite{Mueller2005}}) lessens the amplitude of the surface temperature curve, while smaller solar distances and lower albedos shift the curve towards higher values.

For a first assessment of the temperature range inside the lander, a two-node model has been developed. Using the symmetry of PANIC, the lander can be modeled using two thermal diffusion nodes: One node representing one third of the probe's body, and the other node one of the petals. The lander has been modeled with all petals open during day and night. This means that each node does exchange heat by radiative transfer with the Sun, outer space and the asteroid's surface, while both nodes also exchange heat by radiation with each other. Moreover, some conductive heat flow will be caused by the direct contact of the petal and body with the regolith, although an exact calculation is not feasible due to the unknown contact resistance to the surface. The model includes analytically derived view factors and the incidence angle of solar radiation. The internally dissipated heat equals an average power estimate derived from the power budget as indicated earlier in Table~\ref{tab:Power}, neglecting short power peaks.

The analysis showed that the lander should be isolated at its bottom side against heat exchange with the ground to lessen the influence of the large temperature changes during one asteroid day. Investigations revealed that an insulation material with a conductance of 0.02\,W\,m\textsuperscript{-1}\,K\textsuperscript{-1} appears realistic. Using such a material, the calculated temperature of the lander as shown in Figure~\ref{fig:thermal} varies between about -50\,\textdegree C and +65\,\textdegree C during the period of one asteroid day. Both values are within the temperature range of electronic parts which usually need to be qualified according to MIL standards (typically -55\,\textdegree C to +125\,\textdegree C, see e.g.~\cite{MIL-STD-883F}) and other technical standards used in spacecraft engineering. It should be noted that the calculated temperatures must be interpreted as an average value for the whole lander as the model does not resolve the accommodation of different subsystems. Still, it is expected that the variation in temperature will be less close to the lander's center. Availability of more than 4\,W of heater power would allow to further reduce the temperature gradient for critical components.

\paragraph{Electrical Power Subsystem (EPS)}

Considering the mission requirements of I) highest simplicity of the whole system, II) robustness against a long storage phase during cruise (low self-discharge rate), III) low temperatures during night phases, VI) the low system mass, V) the very limited operational lifetime and VI) the moderate power consumption of the payload and subsystems, primary batteries are the best choice for the lander. They offer an energy density considerably higher than secondary cells, and the mass gain of omitted hardware related to recharge (e.g. solar cells) can be converted into a higher allocated primary cell mass.

Primary cell types which are currently used in space flight and offer the highest energy densities and robustness against wide temperature ranges are lithium-thionyl chloride (LiSOCl\textsubscript{2}, $\approx$\,400\,Wh\,kg\textsuperscript{-1}) and lithium sulfur dioxide (LiSO\textsubscript{2}, \textgreater\,225\,Wh\,kg\textsuperscript{-1}) cells~\cite{Ratnakumar2007}. The recently qualified lithium sulfuryl chloride (LiSO\textsubscript{2}Cl\textsubscript{2}) cell used on ESA's FOTON-M3 mission~\cite{Bennetti2008} offers a higher energy density at temperatures of \mbox{0~--~40\,\textdegree C}, but its capacity drops dramatically at temperatures below \mbox{-10\,\textdegree C}. Considering the expected environment on the surface, the suggested power supply consists of eight lithium-thionyl chloride (LiSOCl\textsubscript{2}) SAFT~\mbox{LSH-20} primary cells, arranged as two strings of four cells each connected in parallel (4s2p~configuration). The selected battery type has an extensive flight record on Sojourner~\cite{Sojourner_Batteries} and Philae~\cite{Debus2003}. The configuration leads to an open circuit voltage of $\approx$\,11.8~--~14\,V depending on temperature, enabling all components either to be supplied directly or after voltage conversion. Considering all losses (self-discharge during cruise, conditioning) and a worst case cell temperature of \mbox{-40\,\textdegree C}, the batteries will still provide a total power of 88~Wh~\cite{SAFT2004}. This value raises up to 181\,Wh at a cell temperature of \mbox{20\,\textdegree C}.

\paragraph{Command and Data Handling (C\&DH)}

The C\&DH subsystem of a specialized lander spacecraft is a mission tailored, customized solution, adapted to the payload requirements and  mission scenario. Approximately 128\,MB mass memory will be required to store and process the acquired data on-board. Based on previous work (e.g.~\cite{Fiethe2007, Bleier2004, Baginski2007}) it appears realistic to assume the C\&DH unit can be realized on two PCBs with a PC/104 form factor, weighing 125\,g each and consuming 2.5\,W of power in operational and 0.75\,W in idle mode.

\paragraph{Communication Subsystem (CS)}

Due to a very short transmission path and the absence of an atmosphere, the communication link is very strong (see Table~\ref{tab:LinkBudget}). This theoretically enables high data rates with great safety margins in terms of bit error probability at a moderate output power. The limiting factor is available miniature hardware. Data rates of previous small body landers did not exceed 16\,kbps (Philae~\cite{Bibring2007a}) or 9.6\,kbps (\mbox{MINERVA}~\cite{Kubota2007}). The CS of the failed Beagle~2 lander was designed to transmit data at a rate up to 128\,kbps~\cite{EuropeanSpaceAgency2004}. 

\begin{table*}[htbp] 
\footnotesize
\centering
\caption{Estimated link budget between lander and orbiter. The nominal case assumes a distance of 10\,km, a boresight RX antenna, polarization losses and a low data rate. Every increase in data rate by factor 2 reduces the carrier-to-noise ratio by 3\,dBHz.} 
\label{tab:LinkBudget}

\begin{tabularx}{\textwidth}{>{\raggedright\arraybackslash}X p{10cm} Y} \toprule
\textbf{Downlink} 					& Frequency (S-Band, 2.1~--~2.5\,GHz)	& 2.4\,Ghz \\ \addlinespace

\textbf{Probe Segment (TX)} 				& Input Power					& 3\,dBW \\ 
\textbf{}						& Efficiency of Transmitter (Including Cable Losses, ...) & 30\,\% \\
							& TX Antenna Gain (Patch Antenna) 		& 0\,dBi~$@$~$\pm$60\textdegree \\ \cmidrule{2-3}
							& Effectively Isotropic Radiated Power (EIRP) 	& -2.2\,dBW \\ \addlinespace

\textbf{Transmission Path}				& Range 			& 10\,km \\
\textbf{}						& Free Space Loss 	& 120\,dB \\ \addlinespace

\textbf{Orbiter (RX)}					& RX Antenna Efficiency						& 60\,\% \\
							& RX Antenna Gain (Assumption: Parabolic Antenna, D = 20\,cm) 	& 11.56\,dB \\
							& Loss (Depointing, Cable, Polarization Mismatch)		& 3.5\,dB \\ \cmidrule{2-3}
							& RX Power 								& -114.2\,dBW \\ \addlinespace
							& Gain to Noise Temperature $\frac{G}{T}$ (Estimated System Noise Temperature: 23.36\,dBK)					&-15.3\,dBK\textsuperscript{-1} \\ \addlinespace
							& Carrier-to-Noise-Density Ratio $\frac{C}{N_0}$			& 91\,dBHz \\ \addlinespace
							& Bit Rate (19.2\,kbps)							& 45.84\,dBHz\\	
							& Assumed Band Width (1.2\,\texttimes\,Bit Rate)						& 46.64\,dBHz \\ \cmidrule{2-3}
							& Carrier-to-Noise-Ratio $\frac{C}{N}$							& 44.4\,dB \\
							& Ratio of Received Energy per Bit to Noise Density $\frac{E_b}{N_0}$				& 45.2\,dB \\ \bottomrule

\end{tabularx} 
\end{table*}
    
This study assumes very conservative estimates as no high speed data transmission has been demonstrated by a planetary lander in a comparable context yet. The lander shall transmit data using a S-Band transmitter (TX) with a data rate of at least 19.2\,kbps and a TX input power of 2\,W. A low-power UHF receiver (RX) permanently listens for commands relayed from the orbiter. A patch antenna is integrated into each of the three side petals to cover the entire hemisphere in the local horizontal reference frame. For a digitally modulated signal using binary phase-shift keying (BPSK), a bit error probability (BEP) of 10\textsuperscript{-14} equals an energy per bit to noise density ratio $\left(\frac{E_b}{N_0}\right)$ of about 14.6\,dB. This means that an extremely high safety margin of more than 30\,dB remains in the assumed nominal case, making any error correction algorithm unnecessary and leaving room for optimization to save power (e.g. by decreasing TX output power or transmission time using a higher data rate). An implemented convolutional coding scheme might still be an option to safeguard the transmitted data against bit errors at a much greater distance than 10\,km to the orbiter or other unforeseen difficulties. However, this will double the data volume, increase the required C\&DH processing power and software complexity. References to demonstrated miniature communication subsystems which could be adapted can be found in~\cite{ISIS_CS, Kuhn2007}. The study assumes one PC/104 board for RX and TX each.

\paragraph{Mass Budget}
\label{sec:massbudget}

Table~\ref{tab:MassBudget} gives a detailed summary of PANIC's mass budget. Set margins are 10\,\% for detailed numbers provided by data sheets, publications or through personal communications, 20\,\% for derived numbers from adaptable micro satellite hardware, and 30\,\% when a rough order of magnitude estimate had to be made. The estimated mass of the studied lander using these margin levels is 11.2\,kg. A payload-to-mass ratio of 12\,\% can be derived. The application of an additional 10\,\% system margin results in a total mass envelope of 12.3\,kg. 

\begin{table}[htbp] 
\footnotesize
\centering
\caption{Estimated mass budget of the lander.}

\begin{tabularx}{\columnwidth}{ >{\raggedright\arraybackslash}X C{1.35cm} C{1.65cm} } \toprule
\textbf{Subsystems} & Mass~\lbrack{g}\rbrack & Margin~\lbrack{g}\rbrack\\ \midrule
\multicolumn{3}{l}{\textbf{Payload}} \\
	   \leftskip=0.5cm {incl. Electronics \& Optics}	& 1180 & 199 \\ \addlinespace
\multicolumn{3}{l}{\textbf{Electrical Power Subsystem}}  \\ 
	   \leftskip=0.5cm {Power Control Unit} 	& 220 & 22 \\
	   \leftskip=0.5cm {Converters \& Regulators} 	& 275 & 28 \\
	   \leftskip=0.5cm {Batteries}			& 800 & 80 \\
	   \leftskip=0.5cm {Wires}			& 400 & 40 \\ \addlinespace
\textbf{C\&DH Unit} & 250  & 50 \\ \addlinespace
\multicolumn{3}{l}{\textbf{Communication Subsystem}} \\
	   \leftskip=0.5cm {S-Band TX \& UHF RX}			& 185 & 19 \\
	   \leftskip=0.5cm {Patch Antennas (3 pcs)} 			& 240 & 24 \\ \addlinespace
\multicolumn{3}{l}{\textbf{Structure and Mechanisms}} \\ 
	   \leftskip=0.5cm {Basic Structure}	& 1100 & 330 \\
	   \leftskip=0.5cm {Petals} 				& 1665 & 500 \\
	   \leftskip=0.5cm {Motors \& Gear Heads}		& 65 & 7 \\
	   \leftskip=0.5cm {Hinges / Mechanisms}		& 375 & 113 \\
	   \leftskip=0.5cm {Precision Stage}			& 200 & 100 \\
	   \leftskip=0.5cm {Battery Compartment}		& 300 & 90 \\
	   \leftskip=0.5cm {Mounting Panels}			& 213 & 64 \\
	   \leftskip=0.5cm {Electronic Compartments (Mounting \& Shielding)}		& 1030 & 103 \\
	   \leftskip=0.5cm {Fasteners, Fittings}		& 618 & 62 \\	\addlinespace
\multicolumn{3}{l}{\textbf{Thermal Control Subsystem}} \\
	   \leftskip=0.5cm {Insulation Material} & 197  & 40 \\ \midrule
\textbf{Total Sum} & 9313 & 1871 \\ \midrule
\multicolumn{3}{l}{\textbf{Resulting Mass Estimate}} \\
	   \leftskip=0.5cm {incl. Subsystem Margins} & \multicolumn{2}{c}{11184} \\
 	   \leftskip=0.5cm {plus 10\% System Margin} & \multicolumn{2}{c}{12302}  \\ \bottomrule

\end{tabularx} 
\label{tab:MassBudget}                      
\end{table}                      

\section{Conclusions}
\label{sec:conclusions}

The study demonstrates that significant surface science on an asteroid can be done with a lander of the 10\,kg class. The presented design will be limited to a non-destructive surface analysis without any subsurface access. It will be able to study the bulk geochemical composition with an advanced scientific payload and provide mobility through hopping to investigate surface diversity.

The required orbiter infrastructure and release mechanism have not been studied within this work. The primary spacecraft has to provide a deployment device which safely holds the lander during launch and cruise, keeps it within storage temperature on the interplanetary trajectory, provides power for system checks and instrument calibrations before deployment (to not discharge the lander's primary cells), and releases it with a predefined velocity at the target. Additionally, the orbiter acts as a relay station, transmitting commands from ground operators to the lander as well as science and telemetry data uplinked from the lander to Earth. Hence, it must provide a high gain antenna and transceiver in the selected communication band as well as enough mass memory for intermediate data storage. 

Given the derived lander mass of 12.3\,kg including margins, a total mass envelope of 15\,kg appears realistic to implement the lander with any necessary infrastructure to the main spacecraft. The lander's simple and robust design supports a variety of targets in terms of solar distance, size and taxonomic asteroid class. Hence, after performing some additional studies to analyze better the requirements to the individual carrier spacecraft, PANIC could be proposed as a PI-lead instrument and would be an appealing option for any upcoming NEA mission.

\section*{Acknowledgements}
The authors would like to acknowledge contributions from Ben Rozitis (Open University, Milton Keynes, UK) on providing a C-code of a simplified thermo-physical model allowing estimates of the asteroid's surface temperatures and Julie Bellerose (Carnegie Mellon University, Silicon Valley Campus / NASA ARC, USA) for discussions on landing dynamics. Special thanks for their support throughout the S4P program and beyond go to Erik Asphaug (UC Santa Cruz, USA) and Greg Delory (UC Berkeley, USA) as well as Amy Gilbert-Morley, Natalie Batalha and all other parties at the Systems Teaching Institute (STI) / University Affiliated Research Center (UARC) at NASA Ames who did their utmost to make S4P a success.


\end{document}